\documentclass[12pt]{article}
\usepackage{epsf}
\def\beq{\begin{equation}}
\def\eeq{\end{equation}}
\def\bea{\begin{eqnarray}}
\def\eea{\end{eqnarray}}
\def\bq{\begin{quote}}
\def\eq{\end{quote}}

\def\bq{\begin{quote}}
\def\eq{\end{quote}}

\def\bq{\begin{quote}}
\def\eq{\end{quote}}

\def\gappeq{\mathrel{\rlap {\raise.5ex\hbox{$>$}}
{\lower.5ex\hbox{$\sim$}}}}

\def\lappeq{\mathrel{\rlap{\raise.5ex\hbox{$<$}}
{\lower.5ex\hbox{$\sim$}}}}

\def\bbz{fa Z \kern-8.9pt Z}

\begin{document}

\baselineskip 24pt
\newcommand{\sheptitle}
{Renormalisation Group Analysis of Single Right-handed Neutrino Dominance}

\newcommand{\shepauthor}
{S. F. King and N. Nimai Singh
\footnote{On leave from the Department of Physics,
Gauhati University, Guwahati - 781014, India.}}

\newcommand{\shepaddress}
{Department of Physics and Astronomy,
University of Southampton, Southampton, SO17 1BJ, U.K.}

\newcommand{\shepabstract}
{We perform a renormalisation group (RG)
analysis of neutrino masses and
mixing angles in the see-saw mechanism in the minimal supersymmetric
standard model with three right-handed neutrinos,
including the effects of the heavy neutrino thresholds.
We focus on the case that one of the right-handed neutrinos
provides the dominant contribution to the 23 block of the light
Majorana matrix, causing its determinant to approximately
vanish and giving an automatic neutrino mass hierarchy, so called
single right-handed neutrino dominance which
may arise from a $U(1)$ family symmetry.
In these models radiative corrections can increase
atomospheric and solar neutrino mixing by up to about $10\%$ and $5\%$, 
respectively, and may help to achieve bi-maximal mixing.
Significantly we find that the radiative corrections over the 
heavy neutrino threshold region are 
at least as important as those usually considered from the
lightest right-handed neutrino down to low energies.}

\begin{titlepage}
\begin{flushright}
hep-ph/0006229\\
\end{flushright}
\begin{center}
{\large{\bf \sheptitle}}
\bigskip \\ \shepauthor \\ \mbox{} \\ {\it \shepaddress} \\ \vspace{.5in}
{\bf Abstract} \bigskip \end{center} \setcounter{page}{0}
\shepabstract
\end{titlepage}

\section{Introduction}

The latest atmospheric neutrino results based on 1117 days of data
from Super Kamiokande are still consistent with a standard two neutrino
oscillation $\nu_{\mu}\rightarrow \nu_{\tau}$
with a near maximal mixing angle $\sin^2 2\theta_{23} >0.88$ and
a mass square splitting $\Delta m_{23}^2$ from $1.5\times 10^{-3}$ 
to $5\times 10^{-3}\ eV^2$
at 90\% CL \cite{sobel}. The sterile neutrino oscillation
hypothesis $\nu_{\mu}\rightarrow \nu_{s}$ is
excluded at 99\% CL.
Super Kamiokande is also beginning to provide important clues
concerning the correct solution to the solar neutrino problem. The
latest results from 1117 days of data from Super Kamiokande \cite{suzuki}
sees a one sigma day-night asymmetry, and a flat energy spectrum,
which together disfavour the small mixing angle (SMA) MSW solution
\cite{MSW},
the just-so vacuum oscillation hypothesis \cite{VO}
and the sterile neutrino hypotheses. 
All three possibilities are now excluded at
95\% CL. The results allow much of the large mixing angle (LMA) MSW
region, which now looks like the leading candidate for the
solution to the solar neutrino problem. For example a typical
point in the LMA MSW region is
$\sin^22\theta_{12} \approx 0.75$ and $\Delta m_{12}^2 \approx 2.5\times
10^{-5}\ eV^2$\cite{LMAMSW}.

The see-saw mechanism \cite{seesaw} implies that the
three light neutrino masses arise from some large mass scales
corresponding to the Majorana
masses of some heavy ``right-handed neutrinos'' $N^p_R$
$M^{pq}_{RR}$ ($p,q=1,\cdots ,Z$) whose entries take values
which extend from $\sim 10^{14}$ GeV
down to perhaps several orders of magnitude lower.
The presence of electroweak scale Dirac mass terms $m_{LR}^{ip}$
(a $3 \times Z$ matrix) connecting the
left-handed neutrinos $\nu^i_L$ ($i=1,\ldots 3$)
to the right-handed neutrinos $N^p_R$
then results in a very light see-saw suppressed effective $3\times 3$ Majorana
mass matrix
\beq
m_{LL}=m_{LR}M_{RR}^{-1}m_{LR}^T
\label{seesaw}
\eeq
for the left-handed neutrinos $\nu_L^i$, which are the light physical
degrees of freedom observed by experiment.
If the neutrino masses arise from the see-saw mechanism then
it is natural to assume the existence of a physical
neutrino mass hierarchy $m_{\nu_1}\ll m_{\nu_2} \ll m_{\nu_3}$,
which implies $\Delta m_{23}^2 \approx m_{\nu_3}^2$, and
$\Delta m_{12}^2 \approx m_{\nu_2}^2$,
which fixes $m_{\nu_3} \approx 5.6 \times 10^{-2}eV$,
and (assuming the LMA MSW solution)
$m_{\nu_2} \approx 5.2 \times 10^{-3}eV$, with rather large errors.
Thus $m_{\nu_2}/m_{\nu_3} \sim 0.1$.
In view of such a 23 mass hierarchy
the presence of a large 23 mixing angle looks a bit surprising
at first sight, especially given our experience with small
quark mixing angles. Several explanations have been proposed
\cite{several}, but the simplest idea is that
the contributions to the 23 block of the light effective Majorana matrix
come predominantly from a single right-handed neutrino, 
which causes the 23 subdeterminant to approximately
vanish. This mechanism, called 
single right-handed neutrino dominance (SRHND), was 
proposed in \cite{SK}, and developed for bi-maximal mixing in \cite{SK2}.

In this paper we shall be concerned with the effect of radiative
corrections in models based on SRHND and
$U(1)$ family symmetry \cite{SK}, and in particular
on the case of bi-maximal mixing in these models \cite{SK2} .
We choose these models because analytic estimates suggest
that the 23 hierarchy arises from a physical mechanism which
accounts for the smallness of the
23 subdeterminant, 
and so this hierarchy should be stable under
radiative corrections. Although
the radiative corrections to atmospheric mixing are only a modest
$10\%$, this can nevertheless play an important role
in achieving near maximal atmospheric mixing.
Our results may be compared to the
several RG studies of various models which 
already exist in the literature
\cite{RG1}, \cite{RG2}. Many of
the existing studies do not take proper account
of the heavy right-handed neutrino mass thresholds,
often only including the effects of running due to the 
low energy dimension 5 see-saw operator from the see-saw scale
down to low energies \cite{RG1}, or running from the high energy scale
down to low energies including a single right-handed neutrino 
mass threshold \cite{RG2}, assuming that all the right-handed
neutrinos are degenerate. Our results indicate that the corrections
in running from the high energy scale, through the heavy neutrino
threshold region, down to the
lightest right-handed neutrino mass scale are just as important
(and in some cases more so) than the traditionally calculated
radiative corrections in running from the lightest right-handed
neutrino mass scale down to low energies.

The paper is organised as follows. In section 2 we define the MSSM
with $Z$ right-handed neutrinos, where the heavy neutrino mass
matrix arises from the vacuum expectation value (VEV) of a
singlet field $\Sigma$, and write
down the renormalisation group equations (RGEs) relevant for
running from the unification scale $M_U \sim 2\times10^{16} GeV$
down to low energies.
In section 3 we review the analytic discussion of the phenomenological
conditions for SRHND and LMA MSW involving
the different types of heavy Majorana neutrino textures,
and show how a $U(1)$ family symmetry may be used to satisfy them \cite{SK2}. 
In section 4 we discuss three explicit examples of this kind,
and then perform our numerical RG analysis of these cases.
Section 5 concludes the paper.

\section{The MSSM with $Z$ Right-handed neutrinos}

We consider the Yukawa terms with two Higgs doublets 
augmented by $Z$ right-handed neutrinos, 
which, are given by
\bea
{\cal L}_{yuk} & = &  \epsilon_{ab}
[ -Y^u_{ij}H_u^aQ_i^bU^c_j
+Y^d_{ij}H_d^aQ_i^bD^c_j
+Y^e_{ij}H_d^aL_i^bE^c_j
-Y^{\nu}_{ip}H_u^aL_i^bN^c_p \nonumber \\
& + & \frac{1}{2}Y_{RR}^{pq}\Sigma N^c_pN^c_q ] + H.c. 
\label{MSSM}
\eea
where $\epsilon_{ab}=-\epsilon_{ba}$, $\epsilon_{12}=1$,
and the remaining notation is standard except that
the $Z$ right-handed neutrinos $N_R^p$ have been replaced by their
CP conjugates $N^c_p$ with $p,q=1,\dots, Z$
and we have introduced a singlet field $\Sigma$ whose
vacuum expectation value (VEV) induces a heavy Majorana matrix 
$M_{RR}=<\Sigma>Y_{RR}$. 
When the two Higgs doublets get their 
VEVS $<H_u^2>=v_2$, $<H_d^1>=v_1$ with $\tan \beta \equiv v_2 /v_1$
we find the terms
\beq
{\cal L}_{yuk}=
v_2Y^u_{ij}U_iU^c_j
+v_1Y^d_{ij}D_iD^c_j
+v_1Y^e_{ij}E_iE^c_j
+v_2Y^{\nu}_{ip}N_iN^c_p + \frac{1}{2}M_{RR}^{pq}N^c_pN^c_q +H.c.
\eeq
Replacing CP conjugate fields we can write in a matrix notation
\beq
{\cal L}_{yuk}=
\bar{U}_Lv_2Y^uU_R
+\bar{D}_Lv_1Y^dD_R
+\bar{E}_Lv_1Y^eE_R
+\bar{N}_Lv_2Y^{\nu}N_R + \frac{1}{2}N^T_RM_{RR}N_R +H.c.
\eeq
where we have assumed that all the masses and Yukawa couplings are
real and written $Y^\ast =Y$.
The diagonal mass matrices are given by the following unitary transformations
\bea
v_2Y^u_{diag}=V_{uL}v_2Y^uV_{uR}^{\dag}={\rm diag (m_u,m_c,m_t)},\nonumber \\ 
v_1Y^d_{diag}=V_{dL}v_1Y^dV_{dR}^{\dag}={\rm diag(m_d,m_s,m_b)},\nonumber \\ 
v_1Y^e_{diag}=V_{eL}v_1Y^eV_{eR}^{\dag}={\rm diag(m_e,m_{\mu},m_{\tau})},
\nonumber \\ 
M_{RR}^{diag}=\Omega_{RR}M_{RR}\Omega_{RR}^{\dag}
={\rm diag(M_{R1},\ldots ,M_{RZ})}.
\label{diagonal}
\eea

Below the mass of the lightest right-handed neutrino
$M_{R1}$ the right-handed neutrino masses may be integrated out
of the theory, which corresponds to replacing the last two terms
in Eq.\ref{MSSM} by a dimension 5 operator
\beq
-\epsilon_{ab}Y^{\nu}_{ip}H_u^aL_i^bN^c_p 
+ \frac{1}{2}M_{RR}^{pq} N^c_pN^c_q +H.c.
\rightarrow
-\frac{1}{2}\kappa_{ij}
(\epsilon_{ab}H_u^aL_i^b)(\epsilon_{a'b'}H_u^{a'}L_j^{b'})+H.c.
\label{See-Saw}
\eeq
where $\kappa=Y_{\nu}M_{RR}^{-1}Y_{\nu}^T$ is simply related 
to the see-saw mass matrix in Eq.\ref{seesaw}
when the Higgs fields are replaced by their VEVs
\beq
m_{LL}=v_2^2\kappa.
\label{mLL}
\eeq
Having constructed the light Majorana mass matrix it must then
be diagonalised by unitary transformations,
\beq
m_{LL}^{diag}=V_{\nu L}m_{LL}V_{\nu L}^{\dag}
={\rm diag(m_{\nu_1},m_{\nu_2},m_{\nu_3})}.
\eeq
The CKM matrix is given by
\beq
V_{CKM}=V_{uL}V_{dL}^{\dag}
\eeq
The leptonic analogue of the CKM matrix
is the MNS matrix defined as \cite{MNS}
\beq
V_{MNS}=V_{eL}V_{\nu L}^{\dag}.
\eeq
which may be parametrised
by a sequence of three rotations about the 1,2 and 3 axes,
as in the standard CKM parametrisation,
\beq
V_{MNS}=R_{23}R_{13}R_{12}
\label{MNS}
\eeq
where
\beq
R_{23}=\left( \begin{array}{rrr}
1 & 0 & 0   \\
0 & c_{23} & s_{23}   \\
0 & -s_{23} & c_{23}
\end{array}
\right),\
R_{13}=\left( \begin{array}{rrr}
c_{13} & 0 & s_{13}   \\
0 & 1 & 0   \\
-s_{13} & 0 & c_{13}
\end{array}
\right),\
R_{12}=\left( \begin{array}{rrr}
c_{12} & s_{12} & 0   \\
-s_{12} & c_{12} & 0   \\
0 & 0 & 1
\end{array}
\right)
\label{123}
\eeq
where $s_{ij}=\sin \theta_{ij}$, $c_{ij}=\cos \theta_{ij}$, and
$\theta_{ij}$ refer to lepton mixing angles.
Note that we completely ignore CP violating phases in this paper.
From the unitarity conditions of the MNS matrix elements in Eq.11, 
and parametrisation in Eq.12, the mixing angles can be expressed 
in terms of elements of $V_{MNS}$ as 
\beq
S_{sol}=\sin^{2}2\theta_{12}=\frac{4V^{2}_{e2}V^{2}_{e1}}
{(V^{2}_{e2}+V^{2}_{e1})^{2}},  
\label{sol}
\eeq
\beq
S_{at}=\sin^{2}2\theta_{23}=\frac{4V^{2}_{{\mu}3}V^{2}_{{\tau}3}}
{(V^{2}_{{\mu}3}+V^{2}_{{\tau}3})^{2}}
\label{at}
\eeq
In principle these mixing angles may not be exactly the same as those
obtained from two generation analysis of the experimental results.

The renormalisation group equations (RGEs) to one-loop
order are:
\bea
\frac{dY^u}{dt} &=&-\frac{1}{16\pi^2}
[N_q.Y^u+Y^u.N_u+(N_{H_u})Y_u]
\nonumber \\
\frac{dY^d}{dt} &=&-\frac{1}{16\pi^2}
[N_q.Y^d+Y^d.N_d+(N_{H_d})Y_d]
\nonumber \\
\frac{dY^{\nu}}{dt} &=&-\frac{1}{16\pi^2}
[N_l.Y^{\nu}+Y^{\nu}.N_{\nu}+(N_{H_u})Y_{\nu}]
\nonumber \\
\frac{dY^e}{dt} &=&-\frac{1}{16\pi^2}
[N_l.Y^e+Y^e.N_e+(N_{H_d})Y_e]
\nonumber \\
\frac{dY_{RR}}{dt} &=&-\frac{1}{16\pi^2}
[N_{\nu}.Y_{RR}+Y_{RR}.N_{\nu}+(N_{\Sigma})Y_{RR}]
\nonumber \\
\label{YukRGEs}
\eea
where the wavefunction anomalous dimensions are
\begin{eqnarray}
N_q &=&(\frac{8}{3}g_3^2+\frac{3}{2}g_2^2+\frac{1}{30}g_1^2)I
-Y^u{Y^u}^{\dagger}-Y^d{Y^d}^{\dagger}  \nonumber \\
N_u &=&(\frac{8}{3}g_3^2+\frac{8}{15}g_1^2)I
-2{Y^u}^{\dagger}Y^u  \nonumber \\
N_d &=&(\frac{8}{3}g_3^2+\frac{2}{15}g_1^2)I
-2{Y^d}^{\dagger}Y^d  \nonumber \\
N_l &=&(\frac{3}{2}g_2^2+\frac{3}{10}g_1^2)I
-Y^e{Y^e}^{\dagger}-Y^{\nu}{Y^{\nu}}^{\dagger}  \nonumber \\
N_e &=&(\frac{6}{5}g_1^2)I
-2{Y^e}^{\dagger}Y^e \nonumber \\
N_{\nu} &=&
-2{Y^{\nu}}^{\dagger}Y^{\nu}-Y_{RR}^{\dagger}Y_{RR} \nonumber \\
N_{H_u} &=&(\frac{3}{2}g_2^2+\frac{3}{10}g_1^2)
-3Tr({Y^u}^{\dagger}Y^u)-Tr({Y^{\nu}}^{\dagger}Y^{\nu})  \nonumber \\
N_{H_d} &=&(\frac{3}{2}g_2^2+\frac{3}{10}g_1^2)
-3Tr({Y^d}^{\dagger}Y^d)-Tr({Y^e}^{\dagger}Y^e)  \nonumber \\
N_{\Sigma}&=&-Tr(Y_{RR}^{\dagger}Y_{RR})
\end{eqnarray}
where $t=\ln \mu$ ($\mu$ is the $\bar{MS}$ scale)
and $I$ is the unit matrix.
The RGEs for the gauge couplings are
\beq
\frac{dg_i}{dt} =\frac{1}{16\pi^2}b_ig_i^3
\eeq
where $b_i=(\frac{33}{5},1,-3)$.

The RGEs above are used to run the Yukawa matrices down
from high energies (say the unification or string scale)
down to the heaviest right-handed neutrino mass, $M_{RZ}$,
which we assume is equal to the VEV of the $\Sigma$ field.
At this mass scale we perform a rotation of the right-handed neutrino
fields to the basis in which $M_{RR}$ is diagonal, according
to Eq.\ref{diagonal}, so that $Y^{\nu}$ is replaced
by $Y^{'\nu}=Y^{\nu}\Omega_{RR}^{\dagger}$.
In this basis one then runs the 
remaining RGEs (apart from $Y_{RR}$) down through the 
right-handed neutrino thresholds  ${\rm diag(M_{RZ},\ldots ,M_{R1})}$,
decoupling each right-handed neutrino at its mass threshold
from the $Y^{'\nu}$ contributions which appear on the right-hand
side of the RGEs.
To be explicit, we replace on the right-hand sides of the RGEs,
\beq
Y^{'\nu}_{ip}\rightarrow Y^{'\nu}_{ip}\theta_{p}
\eeq
where $\theta_p=\theta (\ln \mu - \ln M_{Rp})$.
For the (diagonal) $Y_{RR}$ on the right-hand side we replace it by
\beq
Y_{RRpp} \rightarrow Y_{RRpp}\theta_{p}.
\eeq
In the next section we shall see that for the cases of interest
it is not necessary to diagonalise $M_{RR}$ in order to 
implement decoupling of the right-handed neutrinos,
and one may remain in the basis defined by the $U(1)$ family charges,
since decoupling is facilitated by the simple structures of $M_{RR}$.

Below the mass of the lightest right-handed neutrino
one may use the RGE for $\kappa$, the
coefficient of the dimension 5 neutrino mass operator,
\beq
\frac{d \kappa}{dt}=-\frac{1}{16 \pi^2}[ 
(6g_2^2+\frac{6}{5}g_1^2)\kappa - 6\kappa Tr(Y^u{Y^u}^{\dagger})
-(Y^e{Y^e}^{\dagger})\kappa - \kappa(Y^e{Y^e}^{\dagger})^T ] 
\label{kappaRGE}
\eeq
In solving Eq.\ref{kappaRGE},
it is convenient to diagonalise $Y^u$ and $Y^e$
at the scale of the lightest right-handed neutrino,
using Eqs.\ref{diagonal}, then make the approximation
of keeping only the largest third family Yukawa eigenvalues.
In the diagonal charged lepton basis, $\kappa$ must be transformed
to $\kappa'$ given by
\beq
\kappa'=V_{eL} \kappa V_{eL}^{\dagger}
\eeq
then, retaining the third family 
$t$ and $\tau$ Yukawa couplings only, the RGEs
for the elements of $\kappa'$ are given by
\beq
\frac{d \kappa'_{ij}}{dt}=-\frac{1}{16 \pi^2}\kappa'_{ij} [ 
6g_2^2+\frac{6}{5}g_1^2 - 6h_t^2
-\delta_{i3}h_{\tau}^2-\delta_{3j}h_{\tau}^2] 
\label{kappa'RGE}
\eeq
Following from Eq.\ref{kappa'RGE} we see that the elements
of $m_{LL}'(M_{R1})=v_2^2\kappa'(M_{R1})$ at high energy are renormalised
down to $m_{LL}'(m_t)=v_2^2\kappa'(m_t)$ at low energy,
ignoring the running of $v_2$, according to
\beq
m_{LL}'(m_t) =e^{\frac{6}{5}I_{g_1}}e^{6I_{g_2}}e^{-6I_t}
\left( \begin{array}{ccc}
m_{LL11}'(M_{R1}) & m_{LL12}'(M_{R1}) & m_{LL13}'(M_{R1})e^{-I_{\tau}}\\
m_{LL21}'(M_{R1}) & m_{LL22}'(M_{R1}) & m_{LL23}'(M_{R1})e^{-I_{\tau}}\\
m_{LL31}'(M_{R1})e^{-I_{\tau}} 
& m_{LL32}'(M_{R1})e^{-I_{\tau}} & m_{LL33}'(M_{R1})e^{-2I_{\tau}}
\end{array}
\right)
\label{mLLren}
\eeq
where 
\beq
I_{f}=\frac{1}{16\pi^2}\int_{\ln m_t}^{\ln M_{R1}}h_{f}^2(t)dt, \ \  
I_{g_{i}}=\frac{1}{16\pi^2}\int_{\ln m_t}^{\ln M_{R1}}g_{i}^2(t)dt
\eeq
where $f=t, \tau$ and $i=1,2,3$.
In running down to $m_t$ the charged lepton matrix will remain
diagonal to good approximation, so that the low energy
MNS matrix is simply given by
\beq
V_{MNS}=V_{\nu L}^{'\dag}.
\eeq
where $V_{\nu L}'$ is the matrix which diagonalises
$m_{LL}'(m_t)$,
\beq
m_{LL}^{'diag}(m_t)=V_{\nu L}'m_{LL}'(m_t)V_{\nu L}^{'\dag}
={\rm diag(m_{\nu_1},m_{\nu_2},m_{\nu_3})}.
\eeq

The  mixing matrix $V_{MNS}$ and hence neutrino mixing angles, 
are the running quantities which can be computed at different 
energy scales. For example, the running of the neutrino mixing angle
relevant to the atmospheric neutrino deficit, $\theta_{23}$, 
can be understood from the evolution equation 
\beq
16\pi^{2}\frac{d}{dt}\sin^{2}2\theta_{23}=-2\sin^{2}2\theta_{23}
(1-\sin^{2}\theta_{23})
(h^{2}_{\tau}-h^{2}_{\mu})\frac{m^{33'}_{LL}+m^{22'}_{LL}}
{m^{33'}_{LL}-m^{22'}_{LL}}
\label{23RGE}
\eeq
This equation (see Babu et al in ref.\cite{several})
describes the evolution of the physical 23 mixing angle
assuming that we are in the diagonal charged lepton mass basis.

\section{Three right-handed neutrinos and SRHND}

We now specialise to three right-handed neutrinos and review the
conditions for achieving SRHND and the LMA MSW solution \cite{SK2}.
The statement of SRHND is that, of the three right-handed neutrinos,
one of them, $N_{R3}$, makes the dominant contribution to the
23 block of $m_{LL}$. This ensures that the 23 sub-determinant 
approximately vanishes, and a 23 mass hierarchy therefore
naturally results. 

We first write the neutrino Yukawa matrix in general as
\beq
Y_{\nu}=
\left( \begin{array}{ccc}
a' & a & d\\
b' & b & e\\
c' & c & f
\end{array}
\right)
\label{N3Dirac}
\eeq
There are now three distinct textures for the heavy Majorana neutrino
matrix which maintain the isolation of the dominant right-handed
neutrino $N_{R3}$, namely the diagonal,
democratic and off-diagonal textures
introduced previously\cite{SK,SK2}.
We consider each of them in turn.
Note that, assuming SRHND,
the contribution to the lepton 23 and 13
mixing angles from the neutrino sector are approximately 
\beq
\tan \theta_{23} \approx \frac{e}{f}, \ \ 
\tan \theta_{13} \approx \frac{d}{\sqrt{e^2+f^2}}, \ \ 
\eeq
so that Super-Kamiokande and CHOOZ \cite{CHOOZ} imply
\beq
d\ll e \approx f 
\label{1st}
\eeq
The condition on the 12 mixing angle such that it is relevant for the LMA MSW
solution is discussed separately for each case below.

\subsection{Diagonal Texture}
\beq
M_{RR}=
\left( \begin{array}{ccc}
X' & 0 & 0    \\
0 & X & 0 \\
0 & 0 & Y
\end{array}
\right)
\label{MRRdiag}
\eeq
We may invert the heavy Majorana matrix and construct
the light Majorana matrix using the see-saw mechanism,
\beq
m_{LL}
=
\left( \begin{array}{ccc}
\frac{d^2}{Y}+\frac{a^2}{X}+\frac{a'^2}{X'}
& \frac{de}{Y} +\frac{ab}{X}+\frac{a'b'}{X'}
& \frac{df}{Y}+\frac{ac}{X}+\frac{a'c'}{X'}    \\
.
& \frac{e^2}{Y} +\frac{b^2}{X}+\frac{b'^2}{X'}
& \frac{ef}{Y} +\frac{bc}{X}+\frac{b'c'}{X'}   \\
.
& .
& \frac{f^2}{Y} +\frac{c^2}{X}+\frac{c'^2}{X'}
\end{array}
\right){v_2^2}
\label{matrix3}
\eeq
The SRHND condition is \cite{SK2}
\beq
\frac{e^2}{Y} \sim \frac{ef}{Y} \sim \frac{f^2}{Y} \gg\frac{xy}{X}, 
\frac{x'y'}{X'} 
\label{SRHNDdiag1}
\eeq
where $x,y \in a,b,c$ and $x',y' \in a',b',c'$.
\footnote{The beauty of SRHND is that it automatically implies
$m_{\nu_2}\ll m_{\nu_3}$ due to the approximately
vanishing 23 subdeterminant, without the need for appeal
to cancellations. To understand this simply drop the
$1/X$ terms and observe that the 23 subdeterminant vanishes
which implies a massless eigenvalue which is a rather extreme
case of a hierarchy!}
The 12 mixing angle determines whether we have
the LMA MSW or SMA MSW solution, and this
depends on the relative magnitude of the {\it sub-dominant} 
entries of $m_{LL}$, as discussed in \cite{SK2}.
The condition for LMA MSW is \cite{SK2}
\beq
max\left(\frac{ab}{X},\frac{ac}{X},\frac{a'b'}{X'},\frac{a'c'}{X'}\right)
\sim
max\left(\frac{b^2}{X},\frac{bc}{X},\frac{c^2}{X}
,\frac{b'^2}{X'},\frac{b'c'}{X'},\frac{c'^2}{X'}\right)
\label{LMAMSWdiag}
\eeq

\subsection{Off-Diagonal Texture}
This is defined by:
\beq
M_{RR}=
\left( \begin{array}{ccc}
0 & X & 0    \\
X & 0 & 0 \\
0 & 0 & Y
\end{array}
\right)
\label{MRRoff-diag}
\eeq
The off-diagonal case is qualitatively different from the
other two cases and gives
\beq
m_{LL}
 =
\left( \begin{array}{ccc}
\frac{d^2}{Y}+\frac{2aa'}{X}
& \frac{de}{Y} +\frac{a'b}{X}+\frac{ab'}{X}
& \frac{df}{Y}+\frac{a'c}{X}+\frac{ac'}{X}    \\
.
& \frac{e^2}{Y} +\frac{2bb'}{X}
& \frac{ef}{Y} +\frac{b'c}{X}+\frac{bc'}{X}   \\
.
& .
& \frac{f^2}{Y} +\frac{2cc'}{X}
\end{array}
\right){v_2^2}
\label{matrix5}
\eeq
SRHND is now defined by the conditions
\beq
\frac{e^2}{Y} \sim \frac{ef}{Y} \sim \frac{f^2}{Y} \gg \frac{xx'}{X} 
\label{SRHNDoffdiag}
\eeq
where $x \in a,b,c$ and $x' \in a',b',c'$,
The LMA MSW solution condition is \cite{SK2}:
\beq
max\left(\frac{a'b}{X},\frac{ab'}{X},\frac{a'c}{X},\frac{ac'}{X}\right)
\sim
max\left(\frac{bb'}{X},\frac{b'c}{X},\frac{bc'}{X},\frac{cc'}{X}\right)
\label{LMAMSWoffdiag}
\eeq

\subsection{Democratic Texture}
The democratic case (assuming the Majorana masses in the upper
block are of the same order but are not exactly equal) is defined by:
\beq
M_{RR}=
\left( \begin{array}{ccc}
X & X & 0    \\
X & X & 0 \\
0 & 0 & Y
\end{array}
\right)
\label{MRRdem}
\eeq
The order of magnitude of $m_{LL}$ is: 
\beq
m_{LL}
 =
\left( \begin{array}{ccc}
\frac{d^2}{Y}+O(\frac{a^2}{X})+O(\frac{a'^2}{X})
& \frac{de}{Y} +O(\frac{ab}{X})+O(\frac{a'b'}{X})
& \frac{df}{Y}+O(\frac{ac}{X})+O(\frac{a'c'}{X})    \\
.
& \frac{e^2}{Y} +O(\frac{b^2}{X})+O(\frac{b'^2}{X})
& \frac{ef}{Y} +O(\frac{bc}{X})+O(\frac{b'c'}{X})   \\
.
& .
& \frac{f^2}{Y} +O(\frac{c^2}{X})+O(\frac{c'^2}{X})
\end{array}
\right){v_2^2}
\label{matrix4}
\eeq
In this case the SRHND conditions are:
\beq
\frac{e^2}{Y} \sim \frac{ef}{Y} \sim \frac{f^2}{Y} \gg \frac{xy}{X} 
\sim \frac{x'y'}{X} 
\label{SRHNDdem}
\eeq
where $x,y \in a,b,c$ and $x',y' \in a',b',c'$.
The LMA MSW condition for a large 12 angle is
\beq
max\left(\frac{ab}{X},\frac{ac}{X},\frac{a'b'}{X},\frac{a'c'}{X}\right)
\sim
max\left(\frac{b^2}{X},\frac{bc}{X},\frac{c^2}{X}
,\frac{b'^2}{X},\frac{b'c'}{X},\frac{c'^2}{X}\right)
\label{LMAMSWdem}
\eeq

\subsection{$U(1)$ Family Symmetry}
Introducing a $U(1)$ family symmetry \cite{FN}, \cite{textures},
\cite{IR}, \cite{Ramond} provides a convenient way to
organise the hierarchies within the various Yukawa matrices.
For definiteness we shall focus on a particular class of model based
on a single pseudo-anomalous $U(1)$ gauged family symmetry \cite{IR}.
We assume that the $U(1)$ is broken by the equal VEVs of two
singlets $\theta , \bar{\theta}$ which have vector-like
charges $\pm 1$ \cite{IR}.
The $U(1)$ breaking scale is set by $<\theta >=<\bar{\theta} >$
where the VEVs arise from a Green-Schwartz mechanism \cite{GS} 
with computable Fayet-Illiopoulos $D$-term which
determines these VEVs to be one or two orders of magnitude
below $M_U$. Additional exotic vector matter with
mass $M_V$ allows the Wolfenstein parameter \cite{Wolf}
to be generated by the ratio \cite{IR}
\beq
\frac{<\theta >}{M_V}=\frac{<\bar{\theta} >}{M_V}= \lambda \approx 0.22
\label{expansion}
\eeq

The idea is that at tree-level the $U(1)$ family symmetry
only permits third family Yukawa couplings (e.g. the top quark
Yukawa coupling). Smaller Yukawa couplings are generated effectively
from higher dimension non-renormalisable operators corresponding
to insertions of $\theta$ and $\bar{\theta}$ fields and hence
to powers of the expansion parameter in Eq.\ref{expansion},
which we have identified with the Wolfenstein parameter.
The number of powers of the expansion parameter is controlled
by the $U(1)$ charge of the particular operator.
The fields relevant to neutrino masses
$L_i$, $N^c_p$, $H_u$, $\Sigma$
are assigned $U(1)$ charges $l_i$, $n_p$, $h_u=0$, 
$\sigma$. From Eqs.\ref{expansion},
the neutrino Yukawa couplings and Majorana mass
terms may then be expanded in powers of the Wolfenstein parameter,
\beq
M_{RR} \sim
\left( \begin{array}{ccc}
\lambda^{|2n_1+\sigma|} & \lambda^{|n_1+n_2+\sigma|} 
& \lambda^{|n_1+n_3+\sigma|}\\
. & \lambda^{|2n_2+\sigma|} & \lambda^{|n_2+n_3+\sigma|} \\
.  & .  & \lambda^{|2n_3+\sigma|} 
\end{array}
\right) <\Sigma >
\label{mRR}
\eeq
The conditions which ensure that the third dominant neutrino
is isolated require that the elements $\lambda^{|n_1+n_3+\sigma|}$,
$\lambda^{|n_2+n_3+\sigma|}$ be sufficiently small.
The diagonal, off-diagonal and democratic textures then emerge
as approximate cases \cite{SK,SK2}.
The neutrino Yukawa matrix is explicitly
\beq
Y_{\nu} \sim
\left( \begin{array}{ccc}
\lambda^{|l_1+n_1|} & \lambda^{|l_1+n_2|} 
& \lambda^{|l_1+n_3|}\\
\lambda^{|l_2+n_1|} & \lambda^{|l_2+n_2|} 
& \lambda^{|l_2+n_3|}\\
\lambda^{|l_3+n_1|} & \lambda^{|l_3+n_2|} 
& \lambda^{|l_3+n_3|}
\end{array}
\right)
\label{Ynu}
\eeq
which may be compared to the notation in Eq.\ref{N3Dirac}.
The requirement
of large 23 mixing and small 13 mixing expressed in Eq.\ref{1st}
becomes
\beq
|n_3+l_2|=|n_3+l_3|, \ \ \ \ |n_3+l_1|-|n_3+l_3|=1 \ or \ 2
\label{modulus}
\eeq
The remaining conditions for the $U(1)$ charges
depend on the specific heavy Majorana
texture under consideration \cite{SK2}.

The charged lepton Yukawa matrix is given by
\beq
Y_{e} \sim
\left( \begin{array}{ccc}
\lambda^{|l_1+e_1|} & \lambda^{|l_1+e_2|} 
& \lambda^{|l_1+e_3|}\\
\lambda^{|l_2+e_1|} & \lambda^{|l_2+e_2|} 
& \lambda^{|l_2+e_3|}\\
\lambda^{|l_3+e_1|} & \lambda^{|l_3+e_2|} 
& \lambda^{|l_3+e_3|}
\end{array}
\right)
\label{Ye}
\eeq
where $e_i$ are the $U(1)$ charges of the charged lepton singlet fields.

For the quarks we shall assume a common form
for the textures of $Y^{u}$ and $Y^{d}$
\beq
Y^{u}\sim\left(\begin{array}{ccc}
            \lambda^{8} & \lambda^{5} & \lambda^{3} \\
            \lambda^{7} & \lambda^{4} & \lambda^{2}  \\
            \lambda^{5} & \lambda^{2} & 1
          \end{array}\right), \ \ \ \ 
Y^{d}\sim\left(\begin{array}{ccc}
           \lambda^{4} & \lambda^{3} & \lambda^{3} \\
           \lambda^{3} & \lambda^{2} & \lambda^{2} \\
           \lambda     &  1          &  1          
         \end{array}\right)\lambda^{n}
\eeq

\section{Renormalisation Group Analysis of SRHND}

In \cite{SK2} we tabulated the simplest charges which satisfy all the 
conditions given above, and so provide a natural account
of the atmospheric and solar neutrinos via the LMA MSW effect.
In Table 1 we consider one example of each of the
three cases, namely case A (Diagonal $M_{RR}$), case B (Off-diagonal
$M_{RR}$) and case C (Democratic $M_{RR}$). The $U(1)$ charges along
with corresponding $M_{RR}$, $Y^{\nu}$, obtained from Eqs.\ref{mRR},
\ref{Ynu},
and other relevant parameters 
are outlined for each case, including the charged lepton charges, are
also shown. Note that the zeroes in $M_{RR}$ appear after small
angle rotations on the right-handed neutrino fields, which will not
affect the perturbative expansion in 
powers of $\lambda$ in $Y^{\nu}$. The order unity
coefficients $b_{ij}$, which are always present in $U(1)$ models,
are defined in this basis.

{\small 
\begin{table}[tbp]
\hfil
\begin{tabular}{|l|l|l|l|} \hline  \hline  
Para- & Case A & Case B & Case C \\ meter & (Diagonal $M_{RR}$) &
                               (Off-diagonal $M_{RR}$) & (Democratic $
                               M_{RR}$) \\ \hline $U(1)$ &
                               $l_{1,2,3}=-1,1,1 $ &
                               $l_{1,2,3}=-2,0,0 $ & $
                               l_{1,2,3}=-1,1,1, $ \\ charges &
                               $n_{1,2,3}=1/2,0,-1/2$ &
                               $n_{1,2,3}=-2,1,-1$ &
                               $n_{1,2,3}=0,0,-1/2$ \\ &
                               $e_{1,2,3}=-3,-3,-1,$ &
                               $e_{1,2,3}=-2,-2,0,$ &
                               $e_{1,2,3}=-3,-3,-1,$ \\ &
                               $\sigma=-1$ & $\sigma=1$ &
                               $\sigma=-1$ \\ \hline $Y^{e}$ &
                               $\left(\begin{array}{ccc}
                               a_{11}\lambda^4 & a_{12}\lambda^4 &
                               a_{13}\lambda^2 \\ a_{21}\lambda^2 &
                               a_{22}\lambda^2 & a_{23} \\
                               a_{31}\lambda^2 & a_{32}\lambda^2 &
                               a_{33} \end{array}\right)$ & $
                               \left(\begin{array}{ccc}
                               a_{11}\lambda^4 & a_{12}\lambda^4 &
                               a_{13}\lambda^2 \\ a_{21}\lambda^2 &
                               a_{22}\lambda^2 & a_{23} \\
                               a_{31}\lambda^2 & a_{32}\lambda^2 &
                               a_{33} \end{array}\right)$ & $
                               \left(\begin{array}{ccc}
                               a_{11}\lambda^4 & a_{12}\lambda^4 &
                               a_{13}\lambda^2 \\ a_{21}\lambda^2 &
                               a_{22}\lambda^2 & a_{23} \\
                               a_{31}\lambda^2 & a_{32}\lambda^2 &
                               a_{33} \end{array}\right)$ \\ \hline
                               $a_{ij}$ & $\left(\begin{array}{ccc}
                               0.9 & 1.25 & 0.85 \\ 1.2 & 4.2 & 1.25
                               \\ 0.85 & 1.6 & 1.0 \end{array}\right)$
                               & $ \left(\begin{array}{ccc} 0.9 & 1.25
                               & 0.85 \\ 1.2 & 4.2 & 1.25 \\ 0.85 &
                               1.6 & 1.0 \end{array}\right)$ & $
                               \left(\begin{array}{ccc} 0.9 & 1.25 &
                               0.85 \\ 1.2 & 4.2 & 1.25 \\ 0.85 & 1.6
                               & 1.0 \end{array}\right)$ \\ \hline
                               $Y^{\nu}$ & $ \left(\begin{array}{ccc}
                               b_{11}\lambda^{\frac{1}{2}} &
                               b_{12}\lambda &
                               b_{13}\lambda^{\frac{3}{2}} \\
                               b_{21}\lambda^{\frac{3}{2}} &
                               b_{22}\lambda &
                               b_{23}\lambda^{\frac{1}{2}} \\
                               b_{31}\lambda^{\frac{3}{2}} &
                               b_{32}\lambda &
                               b_{33}\lambda^{\frac{1}{2}}
                               \end{array}\right)$ & $
                               \left(\begin{array}{ccc}
                               b_{11}\lambda^{4} & b_{12}\lambda &
                               b_{13}\lambda^{3} \\ b_{21}\lambda^{2}
                               & b_{22}\lambda & b_{23}\lambda \\
                               b_{31}\lambda^{2} & b_{32}\lambda &
                               b_{33}\lambda \end{array}\right)$ & $
                               \left(\begin{array}{ccc} b_{11}\lambda
                               & b_{12}\lambda &
                               b_{13}\lambda^{\frac{3}{2}} \\
                               b_{21}\lambda & b_{22}\lambda &
                               b_{23}\lambda^{\frac{1}{2}} \\
                               b_{31}\lambda & b_{32}\lambda &
                               b_{33}\lambda^{\frac{1}{2}}
                               \end{array}\right)$ \\ \hline $b_{ij}$
                               & $ \left(\begin{array}{ccc} 0.5 & 0.85
                               & 1.0 \\ 1.0 & 1.3 & 0.4 \\ 1.1 & 0.4 &
                               1.5 \end{array}\right)$ & $
                               \left(\begin{array}{ccc} 1.0 & 1.8 &
                               1.35 \\ 1.8 & 1.35 & 0.4 \\ 1.0 & 0.4 &
                               1.6 \end{array}\right)$ & $
                               \left(\begin{array}{ccc} 0.5 & 0.85 &
                               1.0 \\ 1.0 & 1.3 & 0.5 \\ 1.1 & 0.5 &
                               1.5 \end{array}\right)$ \\ \hline
                               $\frac{M_{RR}}{<\Sigma>}$ & 
                               $ \left(\begin{array}{ccc} 1
                               & 0 & 0 \\ 0 & \lambda & 0 \\ 0 & 0 &
                               \lambda^{2} \end{array}\right)$
                               & $ \left(\begin{array}{ccc} 0 & 1 & 0
                               \\ 1 & 0 & 0 \\ 0 & 0 & \lambda
                               \end{array}\right)$ & $
                               \left(\begin{array}{ccc} \lambda &
                               c_{12}\lambda & 0 \\ 
                               c_{12}\lambda & \lambda & 0 \\
                               0 & 0 & \lambda^2
                               \end{array}\right)$ \\ \hline
                               $<\Sigma>$ & $22.75\times10^{14}GeV $ &
                               $2.20\times10^{14}GeV$ &
                               $38.20\times10^{14}GeV$ \\ \hline
                               $M_{R_{1}}$ & $1.10\times10^{14}GeV$ &
                               $4.84\times10^{13}GeV$ &
                               $1.85\times10^{14}GeV$ \\ \hline
                               $I_{g_{1,2}}$ & $0.048918,0.0751076$ &
                               $0.046789,0.072753$ &
                               $0.0498926,0.07625$ \\ $I_{t,\tau}$ &
                               $0.133729,0.118012$ &
                               $0.130667,0.111514$ &
                               $0.135464,0.120819$ \\ $I_{\mu,e}$ &
                               $4.1884.10^{-4},1.794.10^{-8}$ &
                               $4.0474.10^{-4},1.711.10^{-8}$ &
                               $4.278.10^{-4},1.829.10^{-8}$ \\ 
                               \hline \hline  

\end{tabular}

\hfil
\caption{\footnotesize Textures of the Yukawa couplings of Dirac
neutrino mass, right-handed Majorana neutrino mass, and also other
relevant parameters (as defined in the text)
required for the numerical estimation of
left-handed Majorana neutrino masses at low energies through see-saw
mechanism in tables 2,3 and 4. $<\Sigma>$ is taken as a free parameter
and $ M_{R_{1}}$ is the lowest threshold scale in $M_{RR}$.  
We take $c_{12}\approx 0.9$.}
\end{table}
}

In case A, before rotating to the diagonal charged lepton mass basis,
we may easily estimate the order of the 
entries in $m_{LL}$ in Eq.\ref{matrix3} 
from the matrices $Y^{\nu}$ and $M_{RR}$ in Table 1,
and hence verify the SRHND conditions
Eq.\ref{SRHNDdiag1}, and the LMA MSW condition Eq.\ref{LMAMSWdiag}.
Using Eq.\ref{matrix3} we find
\beq
m_{LL}
\sim
\left( \begin{array}{ccc}
\lambda+\lambda+\lambda
& 1 +\lambda+\lambda^2
& 1+\lambda+\lambda^2    \\
.
& \frac{1}{\lambda} +\lambda+\lambda^3
& \frac{1}{\lambda} +\lambda+\lambda^3  \\
.
& .
& \frac{1}{\lambda} +\lambda+\lambda^3
\end{array}
\right)\frac{v_2^2}{<\Sigma >}
\label{matrix33}
\eeq
where the first entry in each element corresponds to the $1/Y$ contributions
from $N_{3R}$, which clearly dominate the 23 block by a factor
of $\lambda^2$, and the 12 and 13 elements by a factor of $\lambda$.
The 13 element is smaller than the elements in the 23 block by a
factor of $\lambda$ leading to a 13 CHOOZ angle of this order.
The subdominant entries in the 12,13,22,23 elements are the same
order, leading to a large 12 angle suitable for LMA MSW.

Similarly 
in case B, we can show that
the SRHND condition Eq.\ref{SRHNDoffdiag} and LMA MSW condition 
Eq.\ref{LMAMSWoffdiag} are satisfied.
We estimate the order of the 
entries in $m_{LL}$ in Eq.\ref{matrix5} 
from the matrices $Y^{\nu}$ and $M_{RR}$ in Table 1 as,
\beq
m_{LL}
\sim
\left( \begin{array}{ccc}
\lambda^5+2\lambda^5
& \lambda^3 +\lambda^5+\lambda^3
& \lambda^3+\lambda^5+\lambda^3    \\
.
& \lambda +2\lambda^3
& \lambda +\lambda^3+\lambda^3  \\
.
& .
& \lambda +2\lambda^3
\end{array}
\right)\frac{v_2^2}{<\Sigma >}
\label{matrix55}
\eeq
where the first entry in each element corresponds to the $1/Y$ contributions
coming from the right-handed neutrino $N_{3R}$, and clearly
dominates the 23 block by a factor of $\lambda^2$.
In this case it does not dominate the other elements outside the 23 block.
The 13 element from $N_{3R}$ is suppressed relative to the 23 block elements
by a factor of $\lambda^2$, leading to a CHOOZ angle of this order.
The subdominant entries in the 12,13,22,23 elements are again the same
order, leading to a large 12 angle suitable for LMA MSW.

Finally in case C, we verify that
the SRHND condition Eq.\ref{SRHNDdem} and LMA MSW condition 
Eq.\ref{LMAMSWdem} are satisfied, by constructing the
entries in $m_{LL}$ in Eq.\ref{matrix4} 
from the matrices $Y^{\nu}$ and $M_{RR}$ in Table 1,
\beq
m_{LL}
\sim
\left( \begin{array}{ccc}
\lambda +\lambda +\lambda
& 1 +\lambda +\lambda
& 1 +\lambda +\lambda    \\
.
& \frac{1}{\lambda} +\lambda +\lambda
& \frac{1}{\lambda} +\lambda +\lambda\\
.
& .
& \frac{1}{\lambda} +\lambda +\lambda
\end{array}
\right)\frac{v_2^2}{<\Sigma >}
\label{matrix44}
\eeq
where the first entry in each element corresponds to the $1/Y$ contributions
coming from the dominant right-handed neutrino, which dominates
in the 23 block by a factor of $\lambda^2$ and in the 12 and 13 elements 
by a factor of  $\lambda$. The 13 element is smaller than the
23 elements by a factor of $\lambda$ leading to a CHOOZ angle of this
order. The subdominant 12,13,22,23 elements are all the same order
leading to LMA MSW.

We now turn to a numerical treatment of these three cases.
In integrating the RGEs down to low energies
the heavy thresholds for the diagonal texture in Eq.\ref{MRRdiag} 
are dealt with exactly as described in the previous section,
except that no rotation is required to get to the diagonal
$M_{RR}$ basis since we already begin with that form.
For the democratic texture in Eq.\ref{MRRdem} there are essentially
only two mass thresholds to consider $X$ and $Y$ since 
$N_{R1}$ and $N_{R2}$ are approximately degenerate with mass $X$ and we can
ignore any small mass difference between them to leading order.
Similarly for the off-diagonal texture in Eq.\ref{MRRoff-diag}
we also have only two mass thresholds to consider $X$ and $Y$ since 
$N_{R1}$ and $N_{R2}$ are now exactly degenerate with mass $X$.
Thus for both democratic and off-diagonal textures
we can replace on the right-hand sides of the RGEs by
\beq
Y^{\nu}_{i1}\rightarrow Y^{\nu}_{i1}\theta_{1}, \ \ 
Y^{\nu}_{i2}\rightarrow Y^{\nu}_{i2}\theta_{2}, \ \ 
Y^{\nu}_{i3}\rightarrow Y^{\nu}_{i3}\theta_{3} 
\eeq
where $\theta_{1,2}=\theta (\ln \mu - \ln X)$,
$\theta_3=\theta (\ln \mu - \ln Y)$.
We replace $Y_{RR}$ on the right-hand side of the RGEs by
\beq
Y_{RRij} \rightarrow Y_{RRij}\theta_{i}\theta_{j}.
\eeq


\begin{table}[tbp]
\hfil
\begin{tabular}{|l|l|} \hline \hline 
      Scale $ \mu=M_{U}=2.0\times10^{16}GeV$ & Scale
$\mu=M_{U}=2.0\times10^{16}GeV$ \\ \hline $Y^{e}=$ & $V_{eL}=$ \\ $
\left(\begin{array}{lll} 2.114.10^{-3} & 2.928.10^{-3} & 4.114.10^{-2}
\\ 5.808.10^{-2} & 0.20328 & 1.250 \\ 4.114.10^{-2} & 7.744.10^{-2} &
1.000 \end{array}\right)$ & $\left(\begin{array}{lll} 0.999 & 0.002 &
-0.044 \\ 0.036 & -0.621 & 0.783 \\ 0.026 & 0.784 & 0.620
\end{array}\right)$
                               
                              \\ 
$Y^{e}_{diag}=diag(4.36.10^{-4},6.61.10^{-2},1.62)$ &
$m_{e}/m_{\mu},m_{\mu}/m_{\tau}=0.0066,0.041$ \\ $m_{LL}^{\prime\nu}=$
& $V_{MNS}=V_{eL}V_{{\nu}L}^{\dagger}=$ \\ $ \left(\begin{array}{lll}
4.192.10^{-3} & 7.724.10^{-3} & 1.534.10^{-2} \\ 7.724.10^{-3} &
5.345.10^{-2} & 6.894.10^{-2} \\ 1.534.10^{-2} & 6.894.10^{-2} &
9.965.10^{-2} \end{array}\right)$ & $ \left(\begin{array}{lll} 0.797 &
-0.593 & 0.115 \\ 0.432 & 0.692 & 0.579 \\ -0.422 & -0.412 & 0.808
\end{array}\right)$

                                    \\ $m_{LL}^{\prime
diag}=diag(2.46\times10^{-4},0.00582,0.151)$ & $
S_{sol}=0.9173,S_{at}=0.8965$ \\ \hline \hline Scale
$\mu=M_{R_{1}}=1.10\times10^{14}GeV$ & Scale
$\mu=M_{R_{1}}=1.10\times10^{14}GeV $ \\ \hline $Y^{e}=$ & $V_{eL}=$
\\ $\left(\begin{array}{lll} 1.411.10^{-3} & 1.358.10^{-3} &
2.621.10^{-2} \\ 4.234.10^{-2} & 0.15298 & 0.90295 \\ 2.890.10^{-2} &
4.766.10^{-2} & 0.71419 \end{array}\right)$ &
$\left(\begin{array}{lll} 0.996 & 0.004 & -0.042 \\ 0.035 & -0.615 &
0.787 \\ 0.023 & 0.788 & 0.615 \end{array}\right)$

                               \\
 $Y^{e}_{diag}=diag(3.74.10^{-4},5.72.10^{-2},1.16)$ &
$m_{e}/m_{\mu},m_{\mu}/m_{\tau}=0.0065,0.049$
                                        
                                    \\
$m_{LL}^{\prime\nu}=$  & $ V_{MNS}=V_{eL}V_{{\nu}L}^{\dagger}=$  \\ 

 $\left(\begin{array}{lll} 3.827.10^{-3} & 6.800.10^{-3} &
1.278.10^{-2} \\ 6.800.10^{-3} & 4.848.10^{-2} & 5.755.10^{-2} \\
1.278.10^{-2} & 5.755.10^{-2} & 7.715.10^{-2} \end{array}\right)$ &
$\left(\begin{array}{lll} 0.790 & -0.602 & 0.118 \\ 0.421 & 0.671 &
0.606 \\ -0.446 & -0.432 & 0.784 \end{array}\right)$ \\
$m_{LL}^{\prime diag}=diag(2.294.10^{-4},0.0054,0.1238)$ &
$S_{sol}=0.92991,S_{at}=0.9366$ \\ \hline \hline Scale $\mu= m_{t}=175
GeV$ & Scale $\mu=m_{t}=175 GeV$ \\ \hline 
$Y^{e}_{diag}=diag(2.89.10^{-4},4.41.10^{-2},0.627)$ &
$m_{e}/m_{\mu},m_{\mu}/m_{\tau}=0.0065,0.0703$

                                     \\ $m_{LL}^{\prime\nu}=$ & $
 V_{MNS}=V_{{\nu}L}^{\dagger}=$ \\ $\left(\begin{array}{lll}
 2.855.10^{-3} & 5.073.10^{-3} & 8.474.10^{-3} \\ 5.073.10^{-3} &
 3.616.10^{-2} & 3.816.10^{-2} \\ 8.474.10^{-3} & 3.816.10^{-2} &
 4.545.10^{-2} \end{array}\right)$ & $\left(\begin{array}{lll} 0.768 &
 -0.628 & 0.124 \\ 0.412 & 0.633 & 0.656 \\ -0.490 & -0.453 & 0.745
 \end{array}\right)$ \\ $m_{LL}^{\prime
 diag}=diag(1.62\times10^{-4},0.003848,0.08046)$ &
 $S_{sol}=0.9606,S_{at}=0.9840$ \\ \hline \hline

\end{tabular}
\hfil
\caption{\footnotesize Results for case A.
Left-handed Majorana neutrino mass matrix in
the diagonal charged lepton basis and the MNS mixing matrix at
different energy scales $M_{U}, M_{R_{1}}, m_{t}$ for the case of
diagonal $M_{RR}$. Neutrino masses are expressed in eV.  }
\end{table}


{\small
\begin{table}[tbp]
\hfil
\begin{tabular}{|l|l|} \hline \hline 
      Scale $ \mu=M_{U}=2.0\times10^{16}GeV$ & Scale
$\mu=M_{U}=2.0\times10^{16}GeV$ \\ \hline $Y^{e}=$ & $V_{eL}=$ \\ $
\left(\begin{array}{lll} 2.114.10^{-3} & 2.928.10^{-3} & 4.114.10^{-2}
\\ 5.808.10^{-2} & 0.20328 & 1.250 \\ 4.114.10^{-2} & 7.744.10^{-2} &
1.000 \end{array}\right)$ & $\left(\begin{array}{lll} 0.999 & 0.002 &
-0.044 \\ 0.036 & -0.621 & 0.783 \\ 0.026 & 0.784 & 0.620
\end{array}\right)$
                               
                              \\ $Y^{e}_{
diag}=diag(4.36.10^{-4},6.61.10^{-2},1.62)$ &
$m_{e}/m_{\mu},m_{\mu}/m_{\tau}=0.0066,0.041$ \\ $m_{LL}^{\prime\nu}=$
& $V_{MNS}=V_{eL}V_{{\nu}L}^{\dagger}=$ \\ $ \left(\begin{array}{lll}
4.457.10^{-5} & -9.965.10^{-4} & 5.172.10^{-3} \\ -9.965.10^{-4} &
3.116.10^{-2} & 3.784.10^{-2} \\ 5.172.10^{-3} & 3.784.10^{-2} &
5.983.10^{-2} \end{array}\right)$ & $ \left(\begin{array}{lll} 0.886 &
-0.462 & 0.042 \\ 0.363 & 0.740 & 0.566 \\ -0.296 & -0.485 & 0.823
\end{array}\right)$

                                    \\ $m_{LL}^{\prime
diag}=diag(-0.21\times10^{-2},0.00701,0.086)$ & $
S_{sol}=0.6731,S_{at}=0.8918$ \\ \hline \hline Scale
$\mu=M_{R_{1}}=4.84\times10^{13}GeV$ & Scale
$\mu=M_{R_{1}}=4.84\times10^{13}GeV $ \\ \hline $Y^{e}=$ & $V_{eL}=$
\\ $\left(\begin{array}{lll} 1.358.10^{-3} & 1.166.10^{-3} &
2.527.10^{-2} \\ 4.072.10^{-2} & 0.148 & 0.868 \\ 2.803.10^{-2} &
4.548.10^{-2} & 0.6940 \end{array}\right)$ & $\left(\begin{array}{lll}
0.999 & 0.005 & -0.043 \\ 0.037 & -0.620 & 0.784 \\ 0.023 & 0.785 &
0.619 \end{array}\right)$

                               \\ $Y^{e}_{
diag}=diag(3.67.10^{-4},5.64.10^{-2},1.12)$ &
$m_{e}/m_{\mu},m_{\mu}/m_{\tau}=0.0065,0.0503$
                                        
                                    \\
$m_{LL}^{\prime\nu}=$  & $ V_{MNS}=V_{eL}V_{{\nu}L}^{\dagger}=$  \\ 

 $\left(\begin{array}{lll} -1.733.10^{-5} & -1.113.10^{-3} &
4.481.10^{-3} \\ -1.113.10^{-3} & 2.972.10^{-2} & 3.333.10^{-2} \\
4.481.10^{-3} & 3.333.10^{-2} & 4.907.10^{-2} \end{array}\right)$ &
$\left(\begin{array}{lll} 0.882 & -0.469 & 0.039 \\ 0.355 & 0.718 &
0.598 \\ -0.309 & -0.514 & 0.800 \end{array}\right)$ \\
$m_{LL}^{\prime diag}=diag(-2.035.10^{-3},0.00659,0.0742)$ &
$S_{sol}=0.6875,S_{at}=0.9201$ \\ \hline \hline
Scale $\mu= m_{t}=175 GeV$
& Scale $\mu=m_{t}=175 GeV$ \\ \hline $Y^{e}_{
diag}=diag(2.85.10^{-4},4.39.10^{-2},0.619)$ &
$m_{e}/m_{\mu},m_{\mu}/m_{\tau}=0.0065,0.0709$

                                     \\ $m_{LL}^{\prime\nu}=$ & $
 V_{MNS}=V_{{\nu}L}^{\dagger}=$ \\ $\left(\begin{array}{lll}
 -1.295.10^{-5} & -8.317.10^{-4} & 2.995.10^{-3} \\ -8.317.10^{-4} &
 2.221.10^{-2} &2.228.10^{-2} \\ 2.995.10^{-3} & 2.228.10^{-2} &
 2.934.10^{-2} \end{array}\right)$ & $\left(\begin{array}{lll} 0.876 &
 0.480 & 0.036 \\ 0.346 & -0.680 & 0.647 \\ -0.335 & 0.555 & 0.762
 \end{array}\right)$ \\ $m_{LL}^{\prime
 diag}=diag(-1.486\times10^{-3},0.00462,0.0484)$ &
 $S_{sol}=0.7102,S_{at}=0.9738$ \\ \hline \hline

\end{tabular}
\hfil
\caption{\footnotesize Results for case B.
Left-handed Majorana neutrino mass matrix in
the diagonal charged lepton basis and the MNS mixing matrix at
different energy scales $M_{U}, M_{R_{1}}, m_{t}$ for the case of
off-diagonal $M_{RR}$. Neutrino masses are expressed in eV.}
\end{table}

\begin{table}[tbp]
\hfil
\begin{tabular}{|l|l|} \hline \hline 
      Scale $ \mu=M_{U}=2.0\times10^{16}GeV$ & Scale
$\mu=M_{U}=2.0\times10^{16}GeV$ \\ \hline $Y^{e}=$ & $V_{eL}=$ \\ $
\left(\begin{array}{lll} 2.114.10^{-3} & 2.928.10^{-3} & 4.114.10^{-2}
\\ 5.808.10^{-2} & 0.20328 & 1.250 \\ 4.114.10^{-2} & 7.744.10^{-2} &
1.000 \end{array}\right)$ & $\left(\begin{array}{lll} 0.999 & 0.002 &
-0.044 \\ 0.036 & -0.621 & 0.783 \\ 0.026 & 0.784 & 0.620
\end{array}\right)$
                               
                              \\ $Y^{e}_{
diag}=diag(4.36.10^{-4},6.61.10^{-2},1.62)$ &
$m_{e}/m_{\mu},m_{\mu}/m_{\tau}=0.0066,0.041$ \\ $m_{LL}^{\prime\nu}=$
& $V_{MNS}=V_{eL}V_{{\nu}L}^{\dagger}=$ \\ $ \left(\begin{array}{lll}
2.751.10^{-3} & 2.878.10^{-3} & 8.715.10^{-3} \\ 2.878.10^{-3} &
3.070.10^{-2} & 4.222.10^{-2} \\ 8.715.10^{-3} & 4.222.10^{-2} &
6.736.10^{-2} \end{array}\right)$ & $ \left(\begin{array}{lll} 0.780 &
-0.619 & 0.095 \\ 0.480 & 0.688 & 0.544 \\ -0.402 & -0.379 & 0.834
\end{array}\right)$

                                    \\ $m_{LL}^{\prime
diag}=diag(2.96\times10^{-5},0.00488,0.096)$ & $
S_{sol}=0.9488,S_{at}=0.8378$ \\ \hline \hline Scale
$\mu=M_{R_{1}}=1.85\times10^{14}GeV$ & Scale
$\mu=M_{R_{1}}=1.85\times10^{14}GeV $ \\ \hline $Y^{e}=$ & $V_{eL}=$
\\ $\left(\begin{array}{lll} 1.391.10^{-3} & 1.274.10^{-3} &
2.572.10^{-2} \\ 4.293.10^{-2} & 0.1550 & 0.9159 \\ 2.935.10^{-2} &
4.864.10^{-2} & 0.7248 \end{array}\right)$ & $\left(\begin{array}{lll}
0.999 & 0.005 & -0.041 \\ 0.035 & -0.616 & 0.787 \\ 0.022 & 0.788 &
0.615 \end{array}\right)$

                               \\ $Y^{e}_{
diag}=diag(3.77.10^{-4},5.77.10^{-2},1.18)$ &
$m_{e}/m_{\mu},m_{\mu}/m_{\tau}=0.0065,0.049$
                                        
                                    \\
$m_{LL}^{\prime\nu}=$  & $ V_{MNS}=V_{eL}V_{{\nu}L}^{\dagger}=$  \\ 

 $\left(\begin{array}{lll} 2.433.10^{-3} & 2.426.10^{-3} &
7.009.10^{-3} \\ 2.426.10^{-3} & 2.753.10^{-2} & 3.471.10^{-2} \\
7.009.10^{-3} & 3.471.10^{-2} & 5.094.10^{-2} \end{array}\right)$ &
$\left(\begin{array}{lll} 0.771 & -0.630 & 0.096 \\ 0.472 & 0.665 &
0.578 \\ -0.428 & -0.401 & 0.810 \end{array}\right)$ \\
$m_{LL}^{\prime diag}=diag(2.674.10^{-5},0.00433,0.07647)$ &
$S_{sol}=0.9604,S_{at}=0.8945$ \\ \hline \hline 
Scale $\mu= m_{t}=175 GeV$
& Scale $\mu=m_{t}=175 GeV$ \\ \hline $Y^{e}_{
diag}=diag(2.89.10^{-4},4.41.10^{-2},0.626)$ &
$m_{e}/m_{\mu},m_{\mu}/m_{\tau}=0.0065,0.0703$

                                     \\ $m_{LL}^{\prime\nu}=$ & $
 V_{MNS}=V_{{\nu}L}^{\dagger}=$ \\ $\left(\begin{array}{lll}
 1.811.10^{-3} & 1.805.10^{-3} & 4.623.10^{-3} \\ 1.805.10^{-3} &
 2.049.10^{-2} & 2.289.10^{-2} \\ 4.623.10^{-3} & 2.289.10^{-2} &
 2.978.10^{-2} \end{array}\right)$ & $\left(\begin{array}{lll} 0.752 &
 -0.652 & 0.099 \\ 0.461 & 0.628 & 0.627 \\ -0.471 & -0.426 & 0.772
 \end{array}\right)$ \\ $m_{LL}^{\prime
 diag}=diag(1.90\times10^{-5},0.00309,0.04897)$ &
 $S_{sol}=0.9798,S_{at}=0.9579$ \\ \hline \hline

\end{tabular}
\hfil
\caption{\footnotesize Results for case C.
Left-handed Majorana neutrino mass matrix in
the diagonal charged lepton basis and the MNS mixing matrix at
different energy scales $M_{U}, M_{R_{1}}, m_{t}$ for the case of
democratic $M_{RR}$. Neutrino masses are expressed in eV.}
\end{table}

}


Tables 2-4 give the numerical values of quantities at three
different energy scales: the GUT scale $M_{U}=2.0\times10^{16}$ GeV,
the lightest right-handed neutrino mass $M_{R_{1}}\sim 10^{14}$ GeV,
and the low energy scale $m_t=175$ GeV.
To begin with the left-handed Majorana neutrino mass matrix 
\footnote{Strictly speaking the see-saw mechanism does not
operate at energy scales higher than the right-handed neutrino
mass scales, but such results do provide a meaningful measure
of the effects of radiative corrections from the GUT scale
down to low energies.} $m_{LL}(M_{X})$
arising from the sum of dominant and subdominant 
contributions from $N_{R3}$ and $N_{R1}$,$N_{R2}$ 
respectively, are numerically computed at the energy scale 
$\mu=M_{U}=2.0\times10^{16}$GeV, as seen in tables 2,3,4
for the three cases (A,B,C) for large $\tan\beta$. The corresponding 
MNS mixing matrices which 
in turn give $S_{sol}$ and  $S_{at}$  along with neutrino masses 
($m_{\nu1}, m_{\nu2}, m_{\nu3}$) are also estimated.
Next, as already discussed 
in section 2, we calculate the radiative corrections
to $m_{LL}'$ and $V_{MNS}$ through the running of the Yukawa couplings
$Y^{u}$, $Y^{d}$, $Y^{\nu}$, $Y^{e}$, $Y_{RR}$, and the gauge
couplings $g_{1}$, $g_{2}$, $g_{3}$ from high energy scale $M_{U}$
through the heavy neutrino threshold region at successive steps down
to the lightest right-handed neutrino mass $(M_{R_{1}})$. The
corresponding $m_{LL}'(M_{R_{1}})$ (in the diagonal charged lepton basis) 
and $V_{MNS}(M_{R_{1}})$ are again
estimated as shown in tables 2,3 and 4 for three cases. 
Further,
the radiative corrections to $m_{LL}'$ and $V_{MNS}$ from the lightest
right-handed neutrino down to low energies ( say top-quark mass scale
$m_{t}$) are taken in the usual way through the running of the
coefficient of the dimension 5 operator $\kappa'$ in the diagonal
charged lepton basis. These low energy results in tables 2,3,4 can
be compared with observational data, and 
$S_{sol},S_{at}$ defined in Eqs.\ref{sol},\ref{at}
and the neutrino masses given as the diagonal elements of
$m_{LL}^{\prime diag}$ are seen to be in good
agreement with atmospheric data and the LMA MSW solutions.
The CHOOZ constraint is also satisfied in all cases,
with the 13 element of $V_{MNS}$ being larger for cases A and C
than for case B, as expected from the analaytical estimates above.
The charged lepton mass ratios are
$m_{e}/m_{\mu}\approx 0.0066$ 
and $m_{\mu}/m_{\tau}\approx 0.07$
at low energy scale,
compared to the experimental values 0.005 and 0.06, respectively. 

It is instructive to examine the RG evolution of the
neutrino masses and mixing angles.
The numerical results in Tables 1-4 
show that neutrino masses $m_{\nu_{2}}$ and
$m_{\nu_{3}}$ are decreasing from high energy scale $M_{U}$ to low
energy scale $m_{t}$ by about $40\%$, but the ratios
$m_{\nu_{2}}/m_{\nu_{3}}$ increase by about
($24\%$, $17\%$, $55\%$) corresponding to low energy ratios
of (0.05, 0.10, 0.08) for cases (A,B,C), respectively.
The atmospheric mixing quantity $S_{at}$ increases by about
$10\%$ in the three cases, and approaches maximal mixing in all cases.
This can be understood from Eq.\ref{23RGE}, which shows that in the
diagonal charged lepton basis, the sign of the RGE evolution is 
determined by the sign of $m_{LL}^{33'}-m_{LL}^{22'}$, which in the
examples in Tables 2,3,4 is always positive. This means that 
the overall sign of the RGE is negative, which implies an increasing
$S_{at}$ as the energy scale is reduced.
Note that the largest contribution to the atmospheric mixing angle
arises from the charged lepton sector, for this choice of parameters,
as is clear from examining $V_{eL}$ in Tables 2,3,4.
The solar mixing quantity $S_{sol}$ also increases, but the increases
of $1-5\%$ are mild in comparison to $S_{at}$.
In cases A,C $S_{sol}$ approaches maximal mixing, while in case B
its low energy value is about 0.71, which
are well near the best fit $sin^{2}2\theta_{12}\approx0.76$.


In Figures 1-3 we display the RG running of the mixing angles 
$S_{sol}$ and  $S_{at}$ and the ratio of neutrino masses
$m_{\nu2}/m_{\nu3}$ as a function of $t=\ln\mu$, for the three cases
A,B,C. We prefer to show the variation in the neutrino mass ratio,
rather than the absolute values of the two neutrino mass
eigenvalues since they will
depend to some extent on the running vacuum expectation value,
although the effect on the 
present analysis is not large since we assume high $\tan \beta$.
We also do not consider the ratio $m_{\nu1}/m_{\nu2}$ since
this is not experimentally measurable in the hierarchical case.

In Figure 1(a) we show the variations of $S_{sol}$ and 
$S_{at}$ with energy scale for case A. As already stated, the
atmospheric mixing angle runs more rapidly than the solar
mixing angle, and although $S_{at}$ starts out smaller 
then $S_{sol}$ at $M_U$, it quickly grows larger.
Note the effect of the heavy right-handed neutrino mass
thresholds which change the slope of the curves,
which grow steeply over the heavy threshold region.
In Figure 1(b) we give the variation of the
neutrino mass ratio $m_{\nu2}/m_{\nu3}$ with $t=\ln\mu$
for case A. Here the effects of the three heavy right-handed
neutrino thresholds is clearly seen, with again a steep rise
in this mass ratio over the heavy threshold region.

In Figure 2(a) we show the variations of $S_{sol}$ and 
$S_{at}$ with energy scale for case B. 
Here the atmospheric angle starts out larger than the
solar angle, but still grows more rapidly.
In Figure 2(b) we give the variation of the
neutrino mass ratio $m_{\nu2}/m_{\nu3}$ with $t=\ln\mu$
for case B. The qualitative shape of this curve is similar
to Figure 1(b), but there are only two heavy neutrino mass
thresholds in this case, and also the mass ratio is larger throughout.

In Figure 3(a) we show the variations of $S_{sol}$ and 
$S_{at}$ with energy scale for case B. 
Here the solar angle starts out much larger than the
atmospheric angle, and as before the atmospheric
angle grows more rapidly and approaches the solar angle.
In Figure 3(b) we give the variation of the
neutrino mass ratio $m_{\nu2}/m_{\nu3}$ with $t=\ln\mu$
for case B. Because of the choice of $c_{12}$ in Table 1,
the two heavy right-handed neutrino mass thresholds are very
close together in this case.

\newpage 

\vbox{
\noindent
\hfil
\vbox{
\epsfxsize=10cm
\epsffile[130 380 510 735]{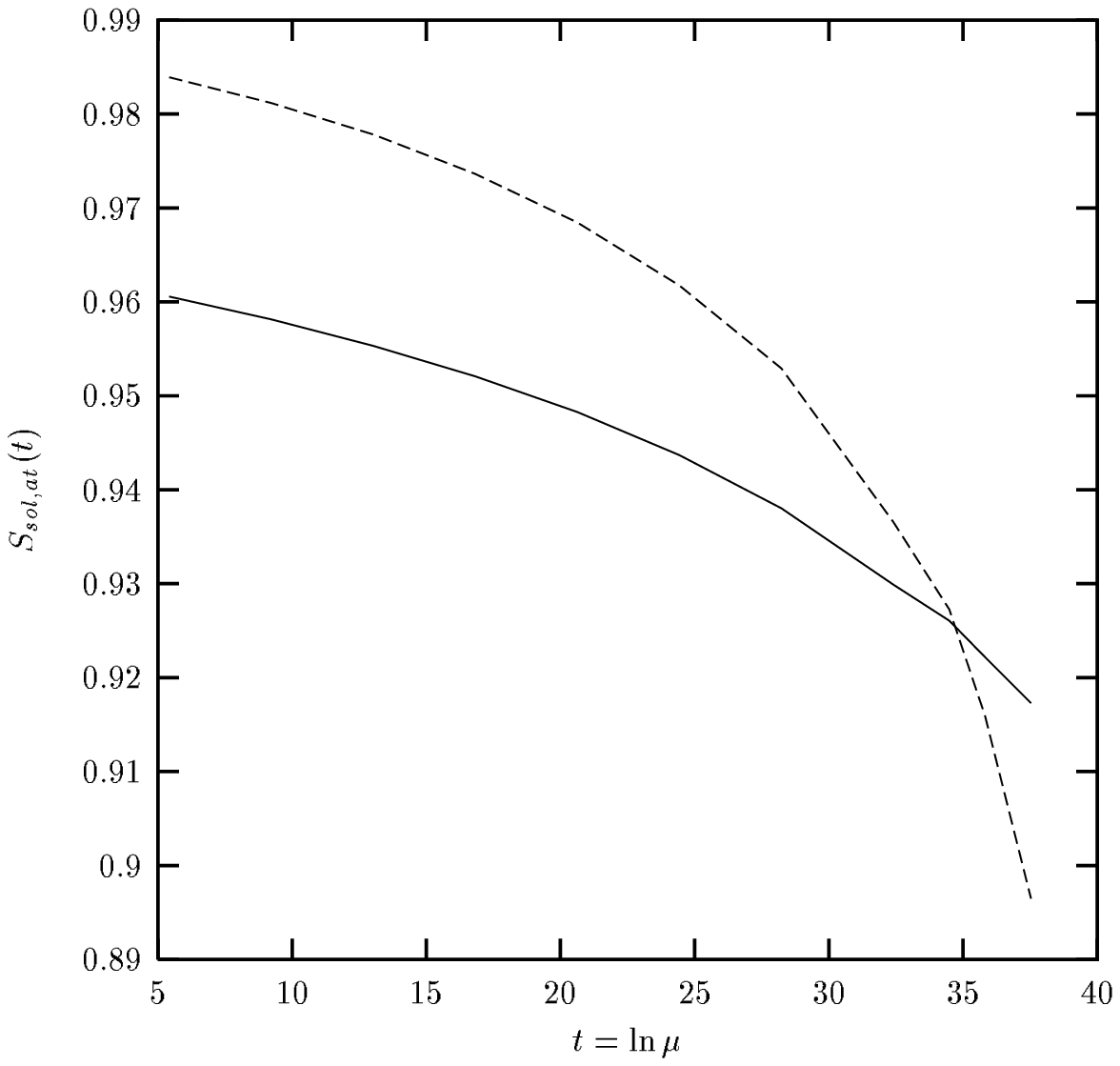}}

{\narrower\narrower\footnotesize\noindent
{Fig.1(a)}
Case A (Diagonal $M_{RR}$): Variation of $S_{sol}$ and 
$S_{at}$ with energy scale
$t=\ln\mu$, which are represented by solid-line and dotted-line
respectively. 
\par}}
\vbox{
\noindent
\hfil
\vbox{
\epsfxsize=10cm
\epsffile[130 380 510 735]{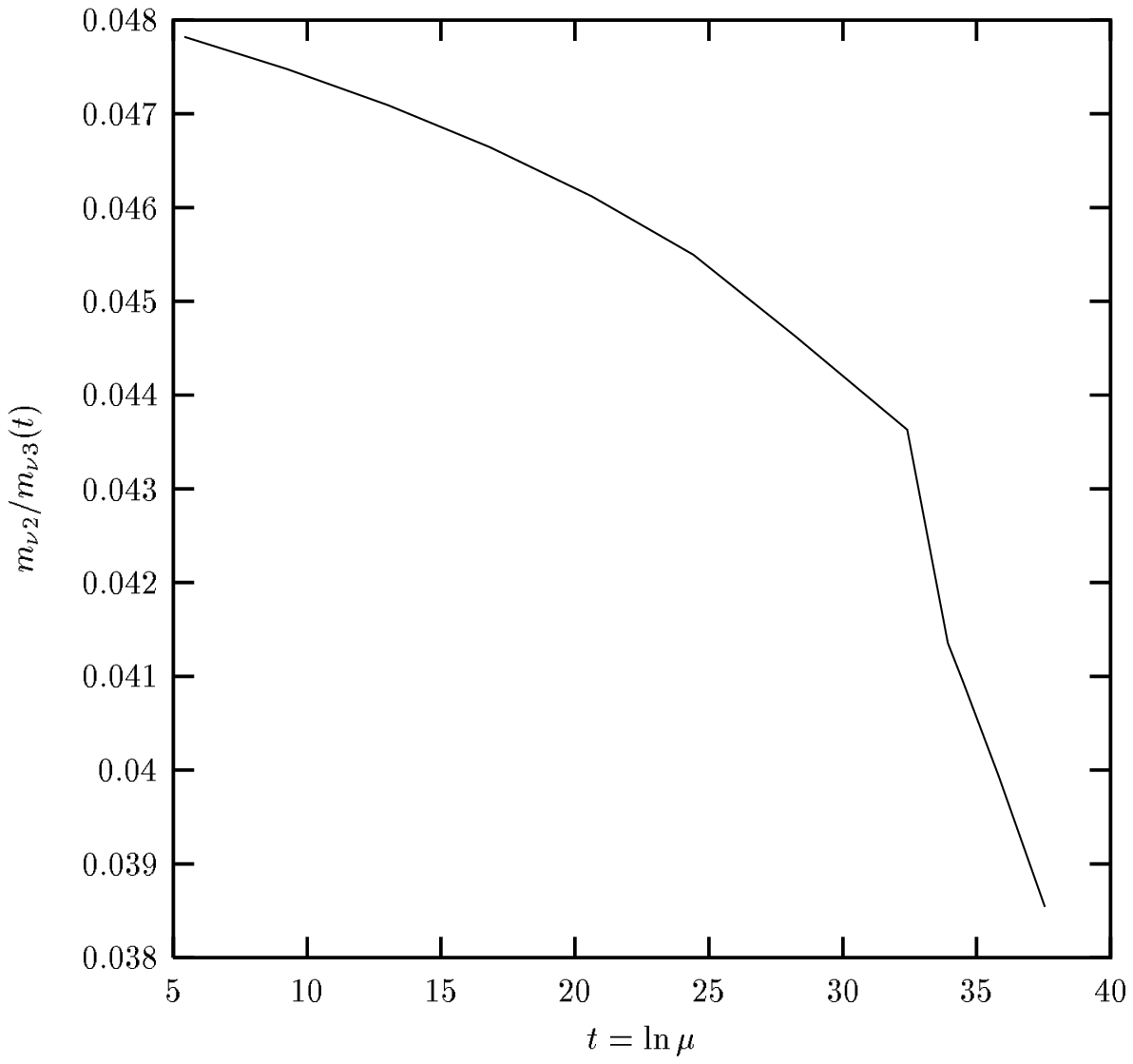}}

{\narrower\narrower\footnotesize\noindent
{Fig.1(b)}
Case A (Diagonal $M_{RR}$): Variation of the 
 neutrino mass ratio $m_{\nu2}/m_{\nu3}$ with $t=\ln\mu$. 
\par}}

\newpage
	
\vbox{
\noindent
\hfil
\vbox{
\epsfxsize=10cm
\epsffile[130 380 510 735]{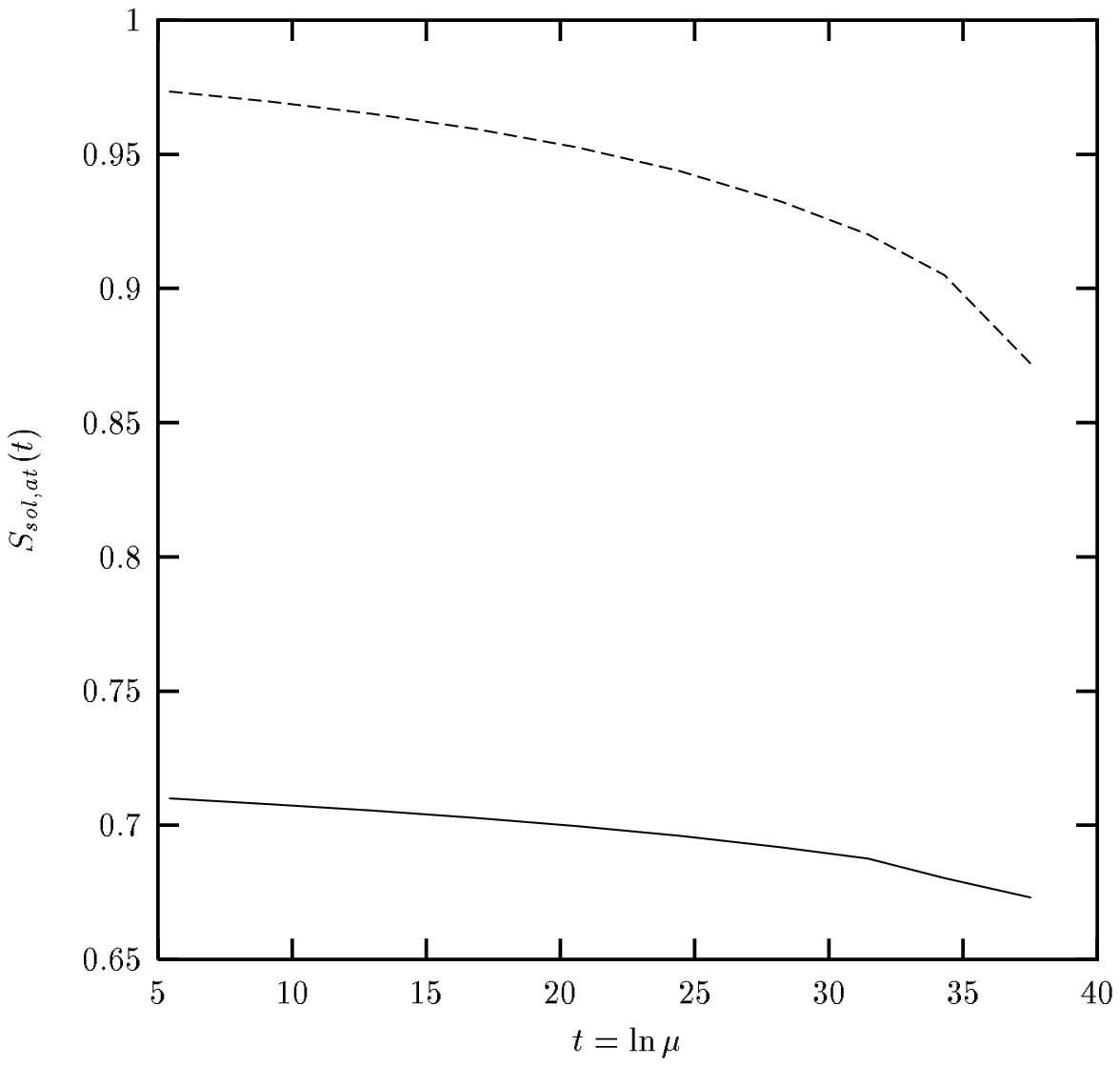}}

{\narrower\narrower\footnotesize\noindent
{Fig.2(a)}
Case B (Off-diagonal $M_{RR}$): Variation of  $S_{sol}$ and 
$S_{at}$ with
$t=\ln\mu$, which are represented by solid-line and dotted-line
respectively. 
\par}}

\vbox{
\noindent
\hfil
\vbox{
\epsfxsize=10cm
\epsffile[130 380 510 735]{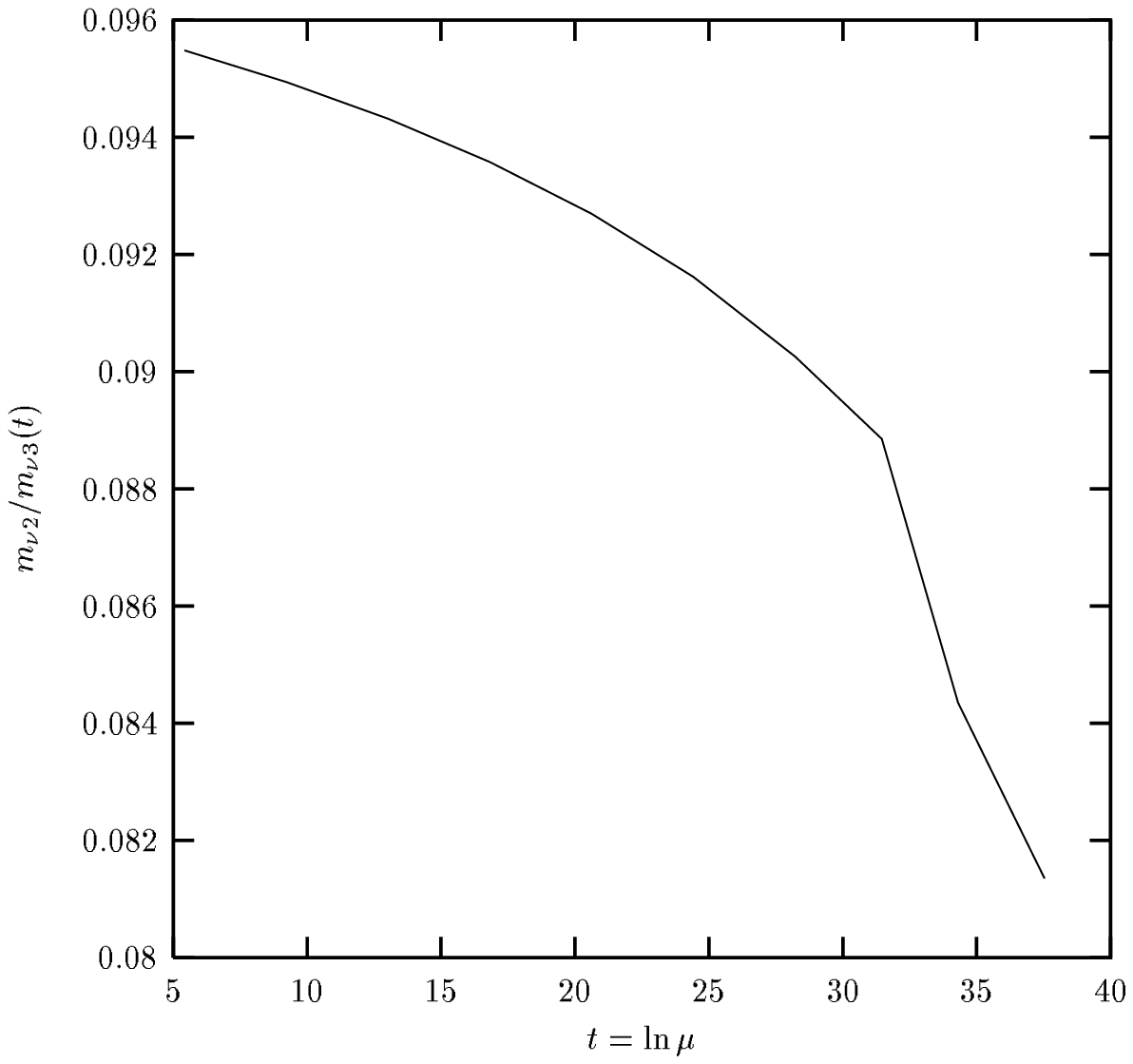}}

{\narrower\narrower\footnotesize\noindent
{Fig.2(b)}
Case B (Off-diagonal $M_{RR}$): Variation of the neutrino mass ratio
  $m_{\nu2}/m_{\nu3}$ with $t=\ln\mu$. 
\par}}

\newpage

\vbox{
\noindent
\hfil
\vbox{
\epsfxsize=10cm
\epsffile[130 380 510 735]{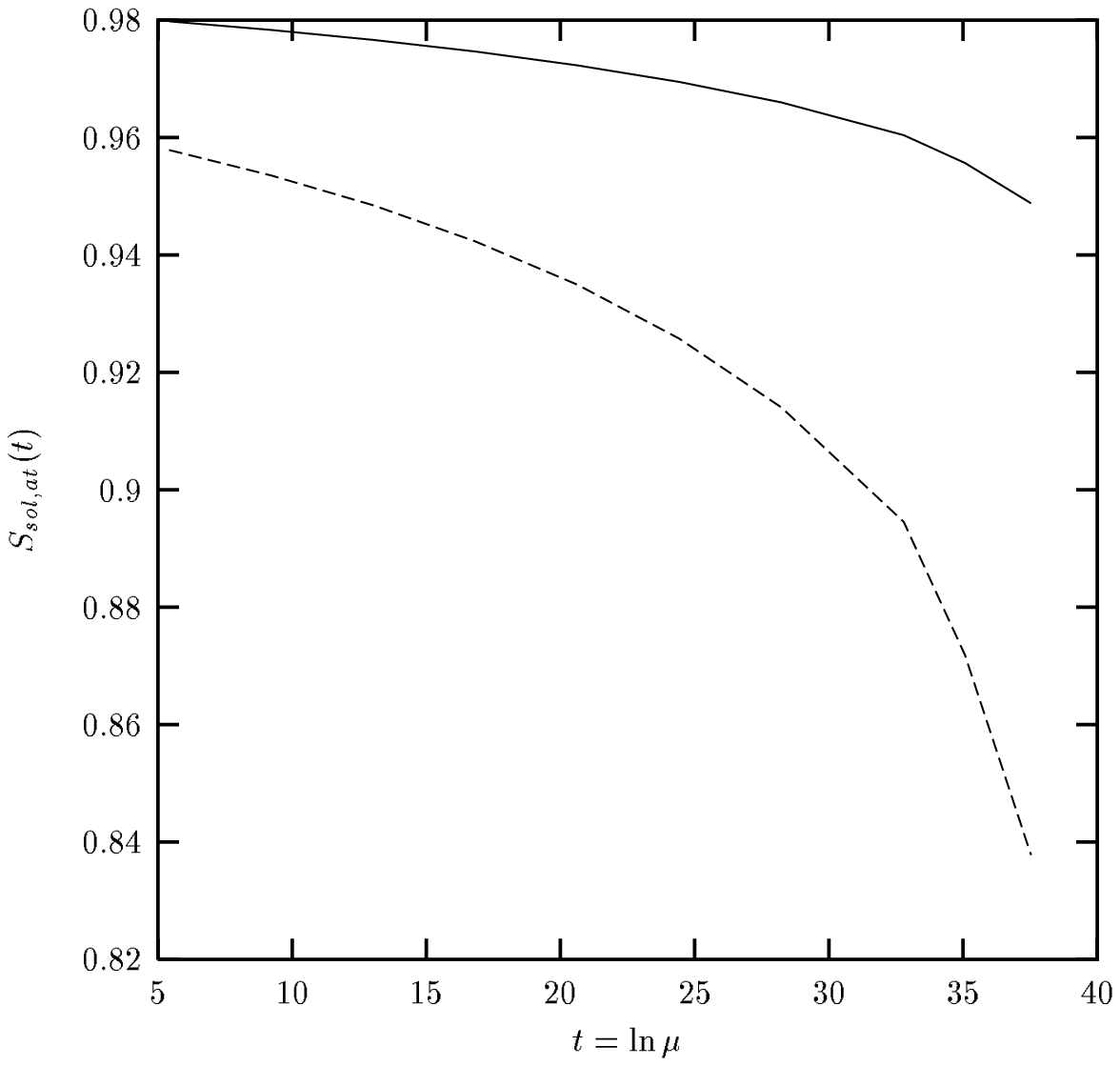}}

{\narrower\narrower\footnotesize\noindent
{Fig.3(a)}
Case C (Democratic $M_{RR}$): Variation of $S_{sol}$ and 
$S_{at}$ with energy scale
$t=\ln\mu$, which are represented by solid-line and dotted-line
respectively. 
\par}}
\vbox{
\noindent
\hfil
\vbox{
\epsfxsize=10cm
\epsffile[130 380 510 735]{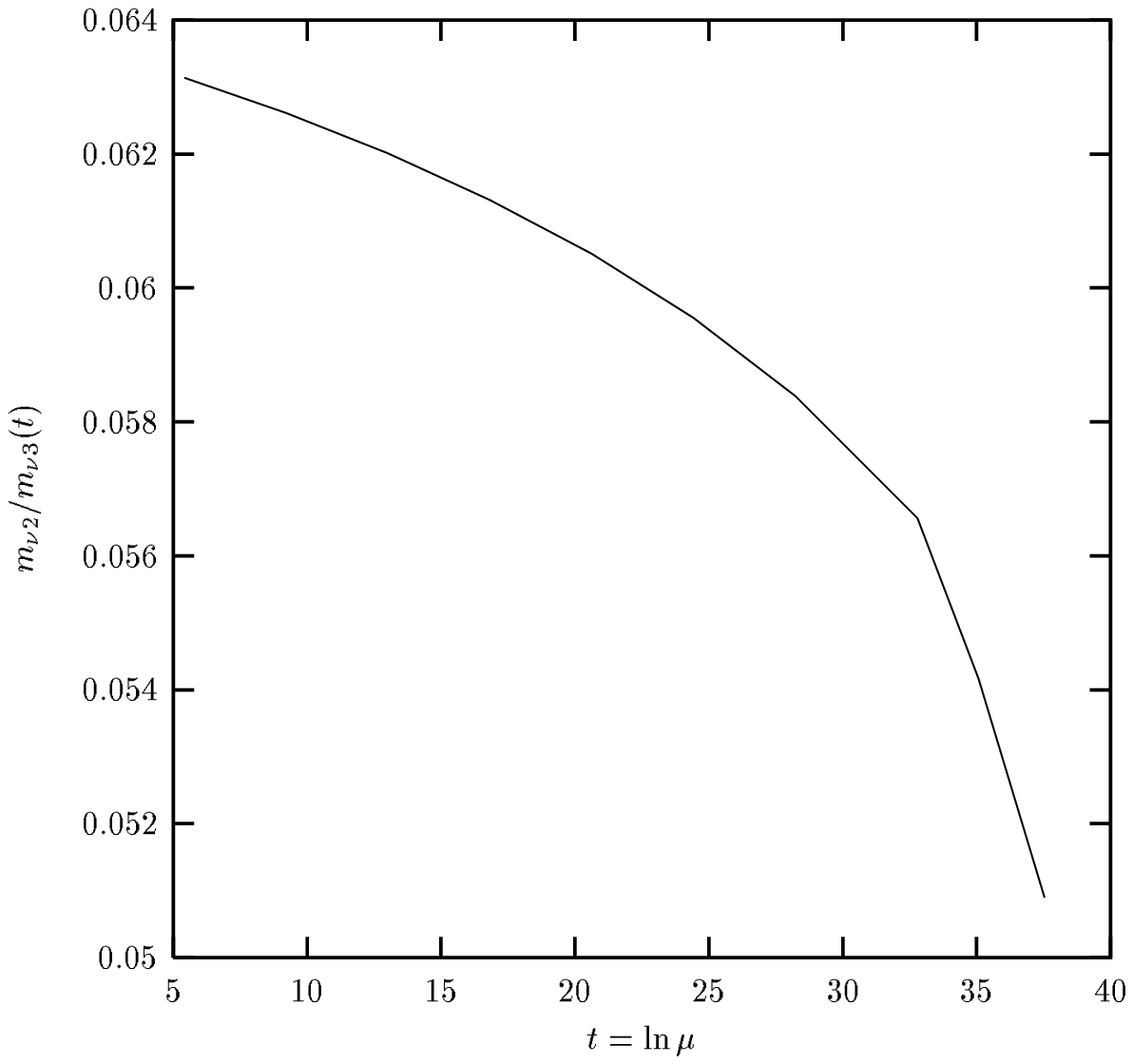}}

{\narrower\narrower\footnotesize\noindent
{Fig.3(b)}
Case C (Democratic $M_{RR}$): Variation of the neutrino mass ratio
 $m_{\nu2}/m_{\nu3}$ with $t=\ln\mu$. 
\par}}


\newpage

It is important to emphasise that more than $50\%$ of the radiative 
corrections to all the physical quantities $m_{\nu_{2}}/m_{\nu_{3}}$,
$S_{at}$, $S_{sol}$ arises from the RG evolution over the range
from the GUT scale $M_{U}=2.0\times10^{16}$GeV through the
heavy right-handed neutrino threshold region down to the lightest
right-handed neutrino mass scale $M_{R_1}\sim 10^{14}$GeV.
Therefore even though this range covers only two orders of magnitude
in energy, it is just as important as the RG evolution effects
from $M_{R_1}\sim 10^{14}$GeV down to low energies which covers
over 12 orders of magnitude in energy, and is the region commonly
considered in the literature.


\section{Conclusion}

We have studied the effects of radiative corrections on neutrino 
masses and mixing angles, in evolving the theory defined at the
GUT scale down to low energies. 
Our approach is to run down all the Yukawa matrices from the
GUT scale down through the heavy right-handed neutrino mass thesholds
to low energy, replacing the neutrino Yukawa matrices by the
dimension 5 neutrino mass operator at the lowest right-handed neutrino
mass threshold. 
We have found that in the realistic cases considered, the
atmospheric and solar neutrino mixing parameters receive
radiative corrections of order
$10\%$ and $5\%$, respectively, 
while the ratios of the neutrino masses change by up about $50\%$
in going from $M_U$ to low energy.
Importantly, more than $50\%$ of the overall radiative corrections
arises from running through the threshold region, even though
it only accounts for two orders of magnitude in energy.
Thus many of the existing analyses in the literature which ignore
this threshold region could endanger a significant error.

We have considered a realistic
class of models known as SRHND \cite{SK,SK2}, in which the contribution
to the 23 block of the light effective Majorana matrix $m_{LL}$
is dominated by a single right-handed neutrino. 
The small neutrino mass hierarchy $m_{\nu_{2}}/m_{\nu_{3}}\sim 0.1$
originates from the smallness of the 23 subdeterminant of $m_{LL}$
(which is zero in the limit that a single right-handed neutrino is
the only contribution). We have focussed on cases with two large
mixing angles $\theta_{23}$ and $\theta_{12}$, with the CHOOZ
angle $\theta_{13}$ being small - the so-called bi-maximal
mixing scenario. The couplings of the dominant right-hand
neutrino controls the 23 and 13 mixing angles, and the subdominant
right-handed neutrino couplings control the 12 angle.
A $U(1)$ family symmetry is used to
generate a controlled expansion of all the Yukawa couplings
in powers of the Wolfenstein parameter
$\lambda$, and with a suitable choice of $U(1)$ charges 
SRHND may be achieved with subdominant right-handed 
neutrino contributions being of the correct order of magnitude
to generate the desired neutrino mixing angles
and mass hierarchy. 

The $U(1)$ charges also control the charged
lepton Yukawa matrix, resulting in large contributions to the
physical lepton mixing angles from the charged lepton sector.
In general one would expect the two contributions to the 23 mixing angle
(from the charged leptons and the neutrinos) not to cancel exactly
due to order unity coefficients in the Yukawa matrices which are
not predicted, and our numerical results include these effects.
Similarly it would be surprising if the physical mixing angles
turned out to be exactly maximal at the GUT scale. Therefore
we have considered examples in which the 23 and 12 angles start out large
but not maximal, and we have also assumed large $\tan \beta$,
where the $\tau$ Yukawa coupling is large and the effect of 
radiative corrections is maximised. Although 
the effect of radiative corrections on the mixing angles
is always $\leq 10\%$,
showing that the models are quite stable, we have shown that the
effects may play an important role in driving the initially
large (but not maximal) mixing angle towards its maximal value.
This is true both of the atmospheric mixing angle
and the solar mixing angle, although in the latter case the
effects are milder, which helps to explain why the
atmospheric angle is larger than the solar angle.
In principle, a different choice of parameters
could have caused the neutrino angles to have grown smaller
at low energies. However it is significant that 
for the cases considered both mixing angles become magnified
showing that
the low energy approximate bi-maximal scenario could partly result from
radiative corrections in SRHND models.

\end{document}